\documentclass[aps,prl,reprint,superscriptaddress]{revtex4-1}

\usepackage{graphicx}
\usepackage{bm}
\usepackage{amsmath}
\usepackage{comment}

\begin{document}


\title{Polarization-dependent interference between dipole moments of a resonantly excited quantum dot}


\author{Disheng Chen}
\affiliation{Department of Physics and Astronomy, West Virginia University, Morgantown, WV 26506, USA}

\author{Gary R. Lander}
\affiliation{Department of Physics and Astronomy, West Virginia University, Morgantown, WV 26506, USA}


\author{Glenn S. Solomon}

\affiliation{Joint Quantum Institute, National Institute of Standards and Technology, \& University of Maryland, Gaithersburg, MD, USA.}

\author{Edward B. Flagg}
\email[]{edward.flagg@mail.wvu.edu}
\affiliation{Department of Physics and Astronomy, West Virginia University, Morgantown, WV 26506, USA}


\date{\today}

\begin{abstract}
Resonant photoluminescence excitation (RPLE) spectra of a neutral InGaAs quantum dot show an unconventional line-shape that depends on the detection polarization.  We characterize this phenomenon by performing polarization-dependent RPLE measurements and simulating the measured spectra with a 3-level quantum model.  This analysis enables us to extract the coherence between the two exciton states from the measured spectra. The good agreement between the data and model indicates that the interference between coherent scatterings from both fine structure split exciton states is the key to understanding this phenomenon.
\end{abstract}


\maketitle


Light-matter interactions in semiconductor nanostructures have attracted significant research interest because of both fundamental physics questions and practical concerns. Epitaxially grown quantum dots (QDs), with their narrow emission linewidths and atom-like density of states in a solid state system, are archetypical elements of study and are potentially useful for many reasons. For example, they have been demonstrated to be efficient sources for single photons \cite{michler_quantum_2000, santori_triggered_2001} and entangled photon pairs \cite{akopian_entangled_2006, trotta_highly_2014}, both of which are capabilities applicable to quantum information science. Resonant excitation of the bound exciton states has allowed measurement of a number of phenomena that cannot be seen with incoherent excitation. Some examples are the Mollow triplet emission from dressed states of a 2-level quantum system \cite{mollow_power_1969, muller_resonance_2007, flagg_resonantly_2009}, and the related Mollow quintuplet from dressed states of a V-system \cite{ge_mollow_2013}. The selectivity and precision of resonant excitation have also allowed the production of high-indistinguishability photons \cite{he_-demand_2013, kuhlmann_transform-limited_2015}, and measurement of charge and spin fluctuations in the local solid state environment \cite{kuhlmann_charge_2013, stanley_dynamics_2014, chen_characterization_2016}.
Many resonant excitation experiments use crossed polarizers to attenuate the laser scattering and allow detection of the resonance fluorescence \cite{vamivakas_spin-resolved_2009, kuhlmann_dark-field_2013, kuhlmann_charge_2013, stanley_dynamics_2014, kuhlmann_transform-limited_2015, he_dynamically_2015}. In such a case, the fluorescence detection is necessarily polarization-selective.
Here we show that when polarization-selective detection is used, orthogonal dipole moments, such as those of a neutral QD or those of a charged QD in a transverse magnetic field, cause an interference effect that results in an unconventional excitation spectrum.

In typical epitaxially grown quantum dots, the anisotropic exchange interaction results in two bound exciton states split by several $\mu$eV with orthogonal transition dipole moments that emit linearly polarized fluorescence \cite{gammon_fine_1996, ivchenko_fine_1997, bayer_fine_2002, tong_theory_2011}.  When the fine structure splitting is on the order of the transition linewidth, a cw excitation laser can interact with both exciton states simultaneously if it is polarized so as to have a non-zero projection onto both dipole moments. Interference between the phase-shifted coherent scattering from the two non-degenerate orthogonal dipoles results in a noticeable difference between the shapes of the excitation spectra for detection polarizations parallel and orthogonal to the excitation. Neither polarized excitation spectrum can be described by the Lorentzian line shapes typically used for 2-level driven quantum systems. By measuring polarization-dependent excitation spectra for polarizations both aligned to the transition dipole moments and 45-degrees rotated relative to them, we can extract the real part of the coherence between the two fine structure states induced by the excitation.

\begin{figure}[b]
	\includegraphics{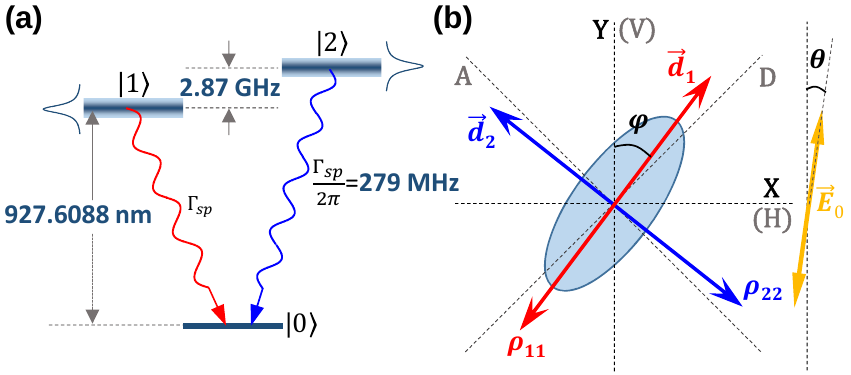}
	\caption{\label{fig:setup}  (a) Quantum dot energy diagram. The population spontaneous decay rate is determined to be $\Gamma_{sp}/2\pi=(279.2\pm 0.9)$ MHz by time-resolved fluorescence measurements (see supplementary information), and the fine structure splitting is $\Delta_\mathrm{FSS}/2\pi=(2.869\pm 0.001)$ GHz.  (b) Polarization and electric dipole moment orientations. The lower energy $\bm{d}_1$ and higher energy $\bm{d}_2$ dipole moments are shown, as is the polarization of the excitation field, $\bm{E}_0$.  The shape of the QD is shown schematically with its asymmetry exaggerated.}
\end{figure}

The sample in this work consists of self-assembled InGaAs QDs embedded in a 4-$\lambda$ GaAs waveguide bounded by two AlGaAs/GaAs distributed Bragg reflectors (DBRs), which form a planar microcavity.  The sample is maintained at 4.2 K in a closed-cycle cryostat.  The QD energy level structure is depicted in Fig.~\ref{fig:setup}(a), and the associated dipole moment orientations are shown in Fig.~\ref{fig:setup}(b). 
\begin{figure*}[!t]
	\includegraphics{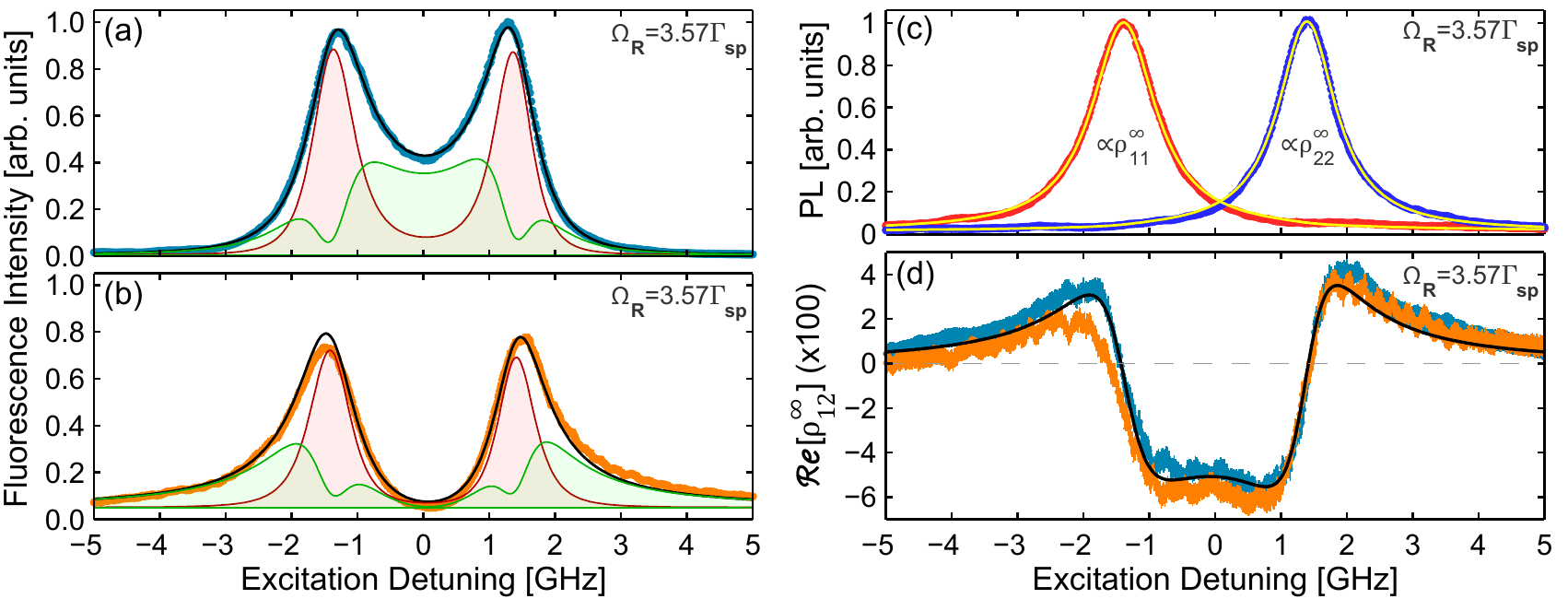}
	\caption{\label{fig:RPLE} Example RPLE spectra recorded at $\Omega_R=3.57 \Gamma_{sp}$ with detection polarization chosen to be (a) X (blue points), (b) Y (orange points), or (c) to eliminate either the low-energy or high-energy peak.  The black solid lines in (a) and (b) are fittings considering the steady state solution for the density matrix.  The green and red shaded regions underneath the spectra are the calculated portion of coherent and incoherent scattering, respectively.  The two yellow curves in (c) are the fittings obtained by using the calculated exciton populations $\rho_{11}^\infty$ and $\rho_{22}^\infty$ from the 3-level model. (d) The coherence $\mathcal{R}e[\rho_{12}^\infty]$ extracted from each curve of (a); the colors indicate the polarization whence the coherence was extracted.  The black solid curve in (d) is not a fit but the calculation of $\mathcal{R}e[\rho_{12}^\infty]$ using the same parameters found in the previous fittings in (a) and (b).}
\end{figure*}
The system is investigated using polarization-dependent resonant photoluminescence excitation (RPLE) spectroscopy, which measures the total fluorescence from the QD as the frequency of the cw excitation laser is scanned across the QD resonance.  Multiple RPLE spectra are recorded using different detection polarizations under the same excitation polarization. Rather than using crossed polarizers to discriminate between fluorescence and laser scattering \cite{vamivakas_spin-resolved_2009, kuhlmann_dark-field_2013}, we instead use modal discrimination between the waveguide mode and the Fabry-Perot mode of the planar microcavity \cite{muller_resonance_2007, flagg_resonantly_2009, chen_characterization_2016, ge_mollow_2013}.  A resonant laser with a 1 MHz linewidth is coupled into the waveguide mode through the cleaved edge of the sample.  The photoluminescence (PL) is coupled out through the Fabry-Perot mode normal to the sample plane. It is first collimated by an aspheric objective lens with NA=0.5, and then it is guided through a pair of liquid crystal variable retarders (LCVRs) and a linear polarizer before being detected by a thermoelectrically cooled CCD camera (see Supplementary Information for a schematic).  The fast axes of the two LCVRs are aligned to the vertical and diagonal directions, allowing us to rotate any polarization state onto the measurement axis determined by the linear polarizer.  Therefore, we can fully characterize the polarization state of the PL via the Stokes vector by measuring the intensity projection on the horizontal (X), vertical (Y), diagonal (D), anti-diagonal (A), left-circular (L), and right-circular (R) polarizations.  

As in previous resonant excitation experiments \cite{chen_characterization_2016, nguyen_photoneutralization_2013}, a small amount of above band-gap excitation is needed to allow resonance fluorescence from the neutral exciton state by neutralizing the intrinsic charges captured by the QD.  The fluorescence of the QD studied in this work is 23 times brighter when introducing 2.19 nW of HeNe laser light (632 nm) onto the sample compared to the case of no above band-gap illumination.  This power corresponds to a fraction of $1.17\times10^{-5}$ times the above band-gap saturation power. 

Between the QD and the LCVRs are several mirrors, which introduce polarization-dependent phase retardance and absorption that change the polarization state of the light.  The effect on the Stokes vector is described by a 4$\times$4 Mueller matrix that can be characterized and measured experimentally \cite{lu_interpretation_1996}.  We characterized the Mueller matrix of the collection path using a laser of the same wavelength as the PL and with multiple polarization states.  The inverse of the resulting Mueller matrix is applied to the measured polarization-dependent data to recover the original light polarization as it was emitted by the QD.  The two principal polarization axes of the PL path, horizontal (X) and vertical (Y), have a transmission coefficient ratio (Y/X) measured to be 0.8067$\pm$0.0598.

Figure~\ref{fig:RPLE}(a)~\&~(b) show two normalized RPLE spectra under the same excitation conditions but with different detection polarizations: horizontal (X) or vertical (Y).  Neither spectrum can be reconstructed by simply adding two Lorentzian lines centered at the two peaks.  We propose that a polarization-dependent interference, occurring between the coherent scattered photons by two neutral exciton states, leads to the observed unconventional line shape.  Clear evidence of the presence of such interference can be seen at zero detuning, where the light scattered by the QD is highly X-polarized, even though the excitation is Y-polarized. 

To prove our hypothesis, we model the neutral QD as a 3-level V-system using a master equation in Lindblad form; we treat the excitation interaction semi-classically under the rotating wave approximation.  The final expressions for the X- and Y-polarized, and total RPLE intensities are as follows (see Supplementary Information for the detailed derivation): 
\begin{gather}
I_X=I_0\{\sin^2(\varphi)\rho_{11}^\infty+\cos^2(\varphi)\rho_{22}^\infty-\sin(2\varphi)\mathcal{R}e[\rho_{12}^\infty]\} \label{eqn:X} \\
I_Y=I_0\{\cos^2(\varphi)\rho_{11}^\infty+\sin^2(\varphi)\rho_{22}^\infty+\sin(2\varphi)\mathcal{R}e[\rho_{12}^\infty]\} \label{eqn:Y} \\
I = I_X + I_Y = I_0\left\{\rho_{11}^\infty+\rho_{22}^\infty\right\} \label{eqn:I}
\end{gather}
where $\rho_{11}^\infty$ and $\rho_{22}^\infty$ are the populations in levels $\left|1\right>$ and $\left|2\right>$, $\rho_{12}^\infty$ is the coherence between the two excited states, the superscript $\infty$ represents the steady state solutions, and $\varphi$ is the angle of the dipole moment $\bm{d_1}$ with respect to the Y-axis (see Fig.~\ref{fig:setup}(b)). $I_0$ is the intensity constant, $I_0={d^2\omega_0^2}/{16\pi^2cr}$, where $d\approx\left|\bm{d}_1\right|\approx\left|\bm{d}_2\right|$ is the magnitude of the transition dipole moment, $\omega_0=(\omega_1+\omega_2)/2$ is the average transition frequency, and $r$ is the distance from the QD to the detector. In this work, we label all the resonant excitation powers with the corresponding overall Rabi frequency $\Omega_R=d\left|\bm{E}_0\right|/\hbar$ in units of the population decay rate $\Gamma_{sp}$.  The individual Rabi frequency $\Omega_j$ for level $\left|j\right>$ is $\Omega_R\cos(\beta_j)$ where $\beta_j$ is the angle between the electric dipole moment $\bm{d}_j$ and the excitation field $\bm{E}_0$.

The RPLE intensities in Eqns.~\ref{eqn:X}~\&~\ref{eqn:Y} are not just proportional to the excited state populations $\rho_{11}^\infty$ and $\rho_{22}^\infty$, which would be the case for a 2-level system. Instead, they are modified by the real part of the coherence between the two excited states, i.e., $\mathcal{R}e[\rho_{12}^\infty]$.  In contrast, the total PL intensity in Eqn.~\ref{eqn:I} is still proportional to the total population in both excited states.  The difference in the sign of the third terms in $I_X$ and $I_Y$ explains the difference between the X-polarized and Y-polarized RPLE spectra.  By simultaneously fitting Eqns.~\ref{eqn:X} and \ref{eqn:Y} to multiple sets of RPLE spectra measured at different excitation powers, we determine the orientation of the dipole moment $\bm{d}_1$ to be $\varphi=44.74^\circ \pm 0.04^\circ$ with respect to the Y-axis, and the direction of the resonant excitation field $\bm{E}_0$ to be $\theta=3.37^\circ \pm 0.07^\circ$ with respect to the Y-axis.  Thus the electric dipole moments of the QD, $\bm{d}_1$ and $\bm{d}_2$, are almost aligned to the diagonal (D) and the anti-diagonal (A) directions, respectively, which is consistent with our Stokes parameter measurement and analysis of the PL (see Supplementary Information).  $\bm{E_0}$ is nominally aligned to the Y-axis because the excitation laser is propagating in the X-direction along the waveguide mode. But $\bm{E_0}$ may deviate from that alignment due to unintentional non-normal incidence of the laser on the air-GaAs interface, which would cause refraction of the beam away from the X-direction.

\begin{figure}[t]
	\includegraphics{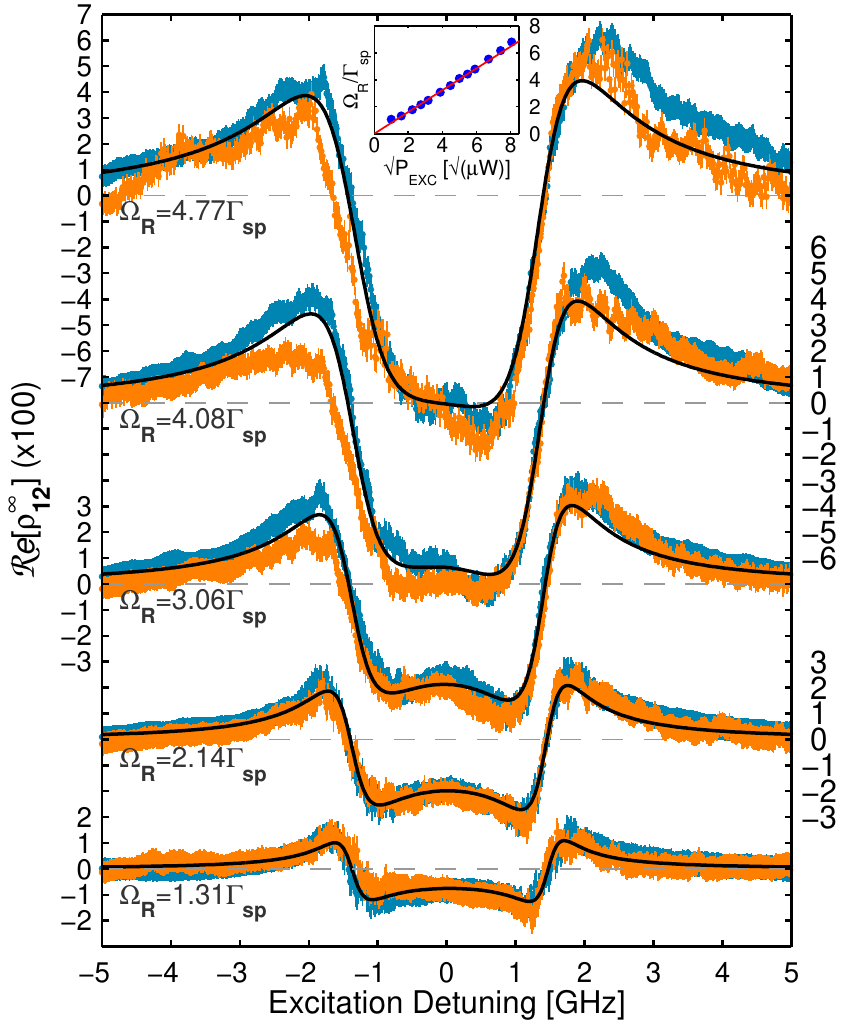}
	\caption{\label{fig:Coherence}  $\mathcal{R}e[\rho_{12}^\infty]$ extracted at different excitation powers.  The curves are vertically offset for clarity.  The color scheme used in each pair is the same as Fig.~\ref{fig:RPLE}(d), i.e., orange is extracted from the Y-polarized RPLE and blue from X-polarized.  The black solid curves are the calculation of $\mathcal{R}e[\rho_{12}^\infty]$ at different Rabi frequencies $\Omega_R$ determined by fittings to the raw RPLE data similar to Figs.~\ref{fig:RPLE}(a-b).  The other model parameters are the same throughout all the calculations.  Inset: Rabi frequency $\Omega_R$ vs.~the square root of the excitation power.  The red straight line is a linear fit with a slope of $(0.806\pm0.011) \Gamma_{sp}/\sqrt{\mu W}$.}
\end{figure} 

Figure~\ref{fig:RPLE}(c) shows the RPLE spectra measured with the LCVRs tuned to block the emission from either the high-energy or low-energy state of the fine structure doublet.  Since that approach measures the emission from only one energy level at a time, there is no interference effect in these spectra. Thus, each spectrum can be directly fitted with the corresponding excited population as $I_1=A_1\rho_{11}^\infty+B_1$  and $I_2=A_2\rho_{22}^\infty+B_2$ for states $\left|1\right>$ and $\left|2\right>$, respectively, where $A_j$ and $B_j$ are fitting parameters.  By doing this fitting, the experimentally measured $\rho_{11}^\infty$ and $\rho_{22}^\infty$ can be found and substituted back into Eqn.~\ref{eqn:X} or \ref{eqn:Y} to solve for the coherence $\mathcal{R}e[\rho_{12}^\infty]$ as shown in Fig.~\ref{fig:RPLE}(d). Note that the sum of the two spectra in Fig.~\ref{fig:RPLE}(c) should be the same as the sum of the previous two spectra in Fig.~\ref{fig:RPLE}(a)~\&~(b) in the sense that both sums are proportional to the total excited population in the QD (Eqn.~\ref{eqn:I}).  However, this equality does not hold for the sum of two raw spectra due to the polarization-dependent absorption of the PL collection path.  Nevertheless, using parameters from the fits in Fig.~\ref{fig:RPLE}(a)~\&~(b), the 3-level V-system simulation reproduces the shape of the coherence successfully (black curve in Fig.~\ref{fig:RPLE}(d)).  

Figure~\ref{fig:Coherence} shows the extracted real part of the coherence for multiple excitation powers. As the excitation power increases, the dispersive line shapes centered at each fine structure resonance increase in magnitude and experience power broadening as is expected for coherent excitation \cite{allen_optical_1975}.  Again, the simulations match the data well and even reproduce the slight asymmetry about zero detuning.  We note that to obtain an observable asymmetry requires two conditions: (1) the dipole moments of the QD must not be oriented 45 degrees with respect to the excitation field, and (2) the excitation power must be high.  The single condition of tilted QD dipole moments is not enough to achieve this asymmetry according to our simulations (see Supplementary Information), implying that this is a non-linear effect happening at high excitation power.  The Rabi frequencies (inset of Fig.~\ref{fig:Coherence}) extracted from the fittings follow a linear relationship with respect to the square root of power, as expected.  

We notice an earlier experimental work \cite{bonadeo_coherent_1998} that utilizes the polarization degree of freedom to study the neutral QD in a pulsed excitation regime.  This is different from \textit{cw} excitation from a fundamental physics point of view.  Under pulsed excitation, a coherent superposition of two orthogonal dipoles is created by the excitation pulse, which then evolves freely over time and experiences quantum beats at a frequency determined by its fine structure splitting $\Delta_{FSS}$.  Since $\Delta_{FSS}\gg\Gamma_{sp}$, it most often leads to a vanishingly small polarization in the time-integrated fluorescence, i.e., zero coherent scattering.  However, in cw excitation we observe a strong polarization, especially at zero detuning where a striking $90^\circ$ polarization switching with respect to the excitation field is present.  An analysis of the proportions of coherent and incoherent scattering in RPLE spectra helps to grasp the underlying physics of this phenomenon.  

In Fig.~\ref{fig:RPLE}(a)~\&~(b), these calculated proportions are denoted by the green and red shaded regions for coherent and incoherent scattering, respectively.  As expected, the incoherent scattering is roughly centered at the two dipole energies, and the slight overlap of the two peaks at zero detuning leads to an ignorable contribution.  In contrast, the coherent scattering is always at the laser wavelength, and a pronounced interference effect is expected between two dipole energies.  In fact, the ``complicated'' envelope of coherent scattering can be understood by noticing: (1) the coherent scattering by a single dipole moment exhibits a dip at its transition energy when the excitation power is not too low, and (2) the relative phase shift of the scattered Rayleigh photons is determined by the detuning with respect to each transition energy.  The latter is similar to a driven harmonic oscillator: red-detuned driving results in a negative lagging phase while blue-detuned driving leads to a positive leading phase.  Finally, the measurement polarization determines how to combine these phase-shifted fields together to either produce constructive interference for horizontal polarization or destructive interference for vertical polarization.  This explains the observed enhancement or diminution of the PL signal around zero detuning in Fig.~\ref{fig:RPLE}(a) and (b), respectively.

\begin{figure}[t]
	\includegraphics{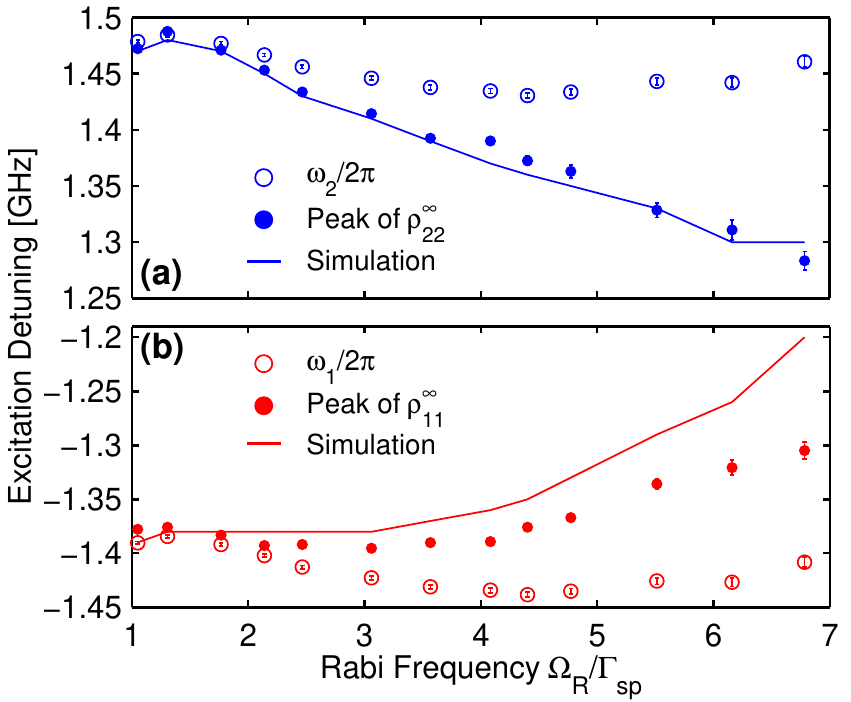}
	\caption{\label{fig:Fit} (a) The evolution of the high-energy state\textsc{\char13}s intrinsic resonance frequency $\omega_2/2\pi$ (blue open circles) and the evolution of the peak positions of the excited population $\rho_{22}^\infty$ obtained by either a Lorentzian fitting (blue dots) or the V-system simulation (blue line).  (b) The evolution of the low-energy state\textsc{\char13}s intrinsic resonance frequency $\omega_1/2\pi$ (red open circles) and the evolution of the peak positions of the excited population $\rho_{11}^\infty$ obtained by either a Lorentzian fitting (red dots) or the V-system simulation (red line).}
\end{figure}

The solid dots in Figs.~\ref{fig:Fit}(a) and (b) are the peak positions of the spectra in Fig.~\ref{fig:RPLE}(c) obtained by fitting them with a Lorentzian function (Fig.~S3 in Supplementary Information).  We find that the two resonance peaks move towards each other as the power increases due to the AC Stark effect.  For example, when the laser is near resonance with the low-energy state $\left|1\right>$, the excitation field is red-detuned with respect to the high-energy state $\left|2\right>$.  
This detuning pushes state $\left|0\right>$ and state $\left|2\right>$ away from each other via the AC Stark effect \cite{unold_optical_2004, muller_emission_2008}.  The red-shift of the ground state in turn effectively increases the transition energy of state $\left|1\right>$; i.e., blue-shift of $\left|1\right>$'s transition energy.  Since the AC Stark effect gets stronger at higher excitation power, the low-energy state $\left|1\right>$ moves continuously towards the higher energy side of the spectrum.  Similarly, the high-energy state $\left|2\right>$ experiences a red-shift in its transition energy as the power increases.  We calculate the resonance positions based on the fitting parameters found in Fig.~\ref{fig:RPLE}(a) and depicted them as solid lines in Fig.~\ref{fig:Fit}(a-b).  They are in good agreement with the data, especially for the high-energy state.

The open circles in Fig.~\ref{fig:Fit}(a) and (b) represent the fitting parameters $\omega_2/2\pi$ and $\omega_1/2\pi$, called the intrinsic transition frequencies for states $\left|2\right>$ and $\left|1\right>$.  These parameters are determined by the intrinsic properties of the QD, which should not change if the QD\textsc{\char13}s local environment is not disturbed.  Therefore, the variation of $\omega_2/2\pi$ and $\omega_1/2\pi$ shown in Figs.~\ref{fig:Fit}(a-b) reflects the fluctuations of the local environment caused by the resonant excitation beam \cite{chen_characterization_2016}.  More discussions on these fluctuations are given in Section 4 of the Supplementary Information.


In this letter, we have demonstrated and modeled an interference effect seen during resonant excitation of a multi-level quantum system.  Similar effects must be accounted for in any situation where there are two non-degenerate orthogonal transition dipole moments and only a certain polarization is detected.  One example is the ``dark-field'' resonant excitation and detection technique \cite{kuhlmann_dark-field_2013} in combination with a charged QD in an in-plane magnetic field.

\begin{acknowledgments}
We would like to acknowledge Tim Thomay for helpful discussions.  This work was supported by the National Science Foundation (DMR-1452840).
\end{acknowledgments}

\bibliography{./Library2}

\begin{thebibliography}{26}%
\makeatletter
\providecommand \@ifxundefined [1]{%
 \@ifx{#1\undefined}
}%
\providecommand \@ifnum [1]{%
 \ifnum #1\expandafter \@firstoftwo
 \else \expandafter \@secondoftwo
 \fi
}%
\providecommand \@ifx [1]{%
 \ifx #1\expandafter \@firstoftwo
 \else \expandafter \@secondoftwo
 \fi
}%
\providecommand \natexlab [1]{#1}%
\providecommand \enquote  [1]{``#1''}%
\providecommand \bibnamefont  [1]{#1}%
\providecommand \bibfnamefont [1]{#1}%
\providecommand \citenamefont [1]{#1}%
\providecommand \href@noop [0]{\@secondoftwo}%
\providecommand \href [0]{\begingroup \@sanitize@url \@href}%
\providecommand \@href[1]{\@@startlink{#1}\@@href}%
\providecommand \@@href[1]{\endgroup#1\@@endlink}%
\providecommand \@sanitize@url [0]{\catcode `\\12\catcode `\$12\catcode
  `\&12\catcode `\#12\catcode `\^12\catcode `\_12\catcode `\%12\relax}%
\providecommand \@@startlink[1]{}%
\providecommand \@@endlink[0]{}%
\providecommand \url  [0]{\begingroup\@sanitize@url \@url }%
\providecommand \@url [1]{\endgroup\@href {#1}{\urlprefix }}%
\providecommand \urlprefix  [0]{URL }%
\providecommand \Eprint [0]{\href }%
\providecommand \doibase [0]{http://dx.doi.org/}%
\providecommand \selectlanguage [0]{\@gobble}%
\providecommand \bibinfo  [0]{\@secondoftwo}%
\providecommand \bibfield  [0]{\@secondoftwo}%
\providecommand \translation [1]{[#1]}%
\providecommand \BibitemOpen [0]{}%
\providecommand \bibitemStop [0]{}%
\providecommand \bibitemNoStop [0]{.\EOS\space}%
\providecommand \EOS [0]{\spacefactor3000\relax}%
\providecommand \BibitemShut  [1]{\csname bibitem#1\endcsname}%
\let\auto@bib@innerbib\@empty
\bibitem [{\citenamefont {Michler}\ \emph {et~al.}(2000)\citenamefont
  {Michler}, \citenamefont {Kiraz}, \citenamefont {Becher}, \citenamefont
  {Schoenfeld}, \citenamefont {Petroff}, \citenamefont {Zhang}, \citenamefont
  {Hu},\ and\ \citenamefont {Imamoglu}}]{michler_quantum_2000}%
  \BibitemOpen
  \bibfield  {author} {\bibinfo {author} {\bibfnamefont {P.}~\bibnamefont
  {Michler}}, \bibinfo {author} {\bibfnamefont {A.}~\bibnamefont {Kiraz}},
  \bibinfo {author} {\bibfnamefont {C.}~\bibnamefont {Becher}}, \bibinfo
  {author} {\bibfnamefont {W.~V.}\ \bibnamefont {Schoenfeld}}, \bibinfo
  {author} {\bibfnamefont {P.~M.}\ \bibnamefont {Petroff}}, \bibinfo {author}
  {\bibfnamefont {L.}~\bibnamefont {Zhang}}, \bibinfo {author} {\bibfnamefont
  {E.}~\bibnamefont {Hu}}, \ and\ \bibinfo {author} {\bibfnamefont
  {A.}~\bibnamefont {Imamoglu}},\ }\href {\doibase
  10.1126/science.290.5500.2282} {\bibfield  {journal} {\bibinfo  {journal}
  {Science}\ }\textbf {\bibinfo {volume} {290}},\ \bibinfo {pages} {2282}
  (\bibinfo {year} {2000})}\BibitemShut {NoStop}%
\bibitem [{\citenamefont {Santori}\ \emph {et~al.}(2001)\citenamefont
  {Santori}, \citenamefont {Pelton}, \citenamefont {Solomon}, \citenamefont
  {Dale},\ and\ \citenamefont {Yamamoto}}]{santori_triggered_2001}%
  \BibitemOpen
  \bibfield  {author} {\bibinfo {author} {\bibfnamefont {C.}~\bibnamefont
  {Santori}}, \bibinfo {author} {\bibfnamefont {M.}~\bibnamefont {Pelton}},
  \bibinfo {author} {\bibfnamefont {G.}~\bibnamefont {Solomon}}, \bibinfo
  {author} {\bibfnamefont {Y.}~\bibnamefont {Dale}}, \ and\ \bibinfo {author}
  {\bibfnamefont {Y.}~\bibnamefont {Yamamoto}},\ }\href {\doibase
  10.1103/PhysRevLett.86.1502} {\bibfield  {journal} {\bibinfo  {journal}
  {Physical Review Letters}\ }\textbf {\bibinfo {volume} {86}},\ \bibinfo
  {pages} {1502} (\bibinfo {year} {2001})}\BibitemShut {NoStop}%
\bibitem [{\citenamefont {Akopian}\ \emph {et~al.}(2006)\citenamefont
  {Akopian}, \citenamefont {Lindner}, \citenamefont {Poem}, \citenamefont
  {Berlatzky}, \citenamefont {Avron}, \citenamefont {Gershoni}, \citenamefont
  {Gerardot},\ and\ \citenamefont {Petroff}}]{akopian_entangled_2006}%
  \BibitemOpen
  \bibfield  {author} {\bibinfo {author} {\bibfnamefont {N.}~\bibnamefont
  {Akopian}}, \bibinfo {author} {\bibfnamefont {N.~H.}\ \bibnamefont
  {Lindner}}, \bibinfo {author} {\bibfnamefont {E.}~\bibnamefont {Poem}},
  \bibinfo {author} {\bibfnamefont {Y.}~\bibnamefont {Berlatzky}}, \bibinfo
  {author} {\bibfnamefont {J.}~\bibnamefont {Avron}}, \bibinfo {author}
  {\bibfnamefont {D.}~\bibnamefont {Gershoni}}, \bibinfo {author}
  {\bibfnamefont {B.~D.}\ \bibnamefont {Gerardot}}, \ and\ \bibinfo {author}
  {\bibfnamefont {P.~M.}\ \bibnamefont {Petroff}},\ }\href {\doibase
  10.1103/PhysRevLett.96.130501} {\bibfield  {journal} {\bibinfo  {journal}
  {Physical Review Letters}\ }\textbf {\bibinfo {volume} {96}},\ \bibinfo
  {pages} {130501} (\bibinfo {year} {2006})}\BibitemShut {NoStop}%
\bibitem [{\citenamefont {Trotta}\ \emph {et~al.}(2014)\citenamefont {Trotta},
  \citenamefont {Wildmann}, \citenamefont {Zallo}, \citenamefont {Schmidt},\
  and\ \citenamefont {Rastelli}}]{trotta_highly_2014}%
  \BibitemOpen
  \bibfield  {author} {\bibinfo {author} {\bibfnamefont {R.}~\bibnamefont
  {Trotta}}, \bibinfo {author} {\bibfnamefont {J.~S.}\ \bibnamefont
  {Wildmann}}, \bibinfo {author} {\bibfnamefont {E.}~\bibnamefont {Zallo}},
  \bibinfo {author} {\bibfnamefont {O.~G.}\ \bibnamefont {Schmidt}}, \ and\
  \bibinfo {author} {\bibfnamefont {A.}~\bibnamefont {Rastelli}},\ }\href
  {\doibase 10.1021/nl500968k} {\bibfield  {journal} {\bibinfo  {journal} {Nano
  Letters}\ }\textbf {\bibinfo {volume} {14}},\ \bibinfo {pages} {3439}
  (\bibinfo {year} {2014})}\BibitemShut {NoStop}%
\bibitem [{\citenamefont {Mollow}(1969)}]{mollow_power_1969}%
  \BibitemOpen
  \bibfield  {author} {\bibinfo {author} {\bibfnamefont {B.~R.}\ \bibnamefont
  {Mollow}},\ }\href {\doibase 10.1103/PhysRev.188.1969} {\bibfield  {journal}
  {\bibinfo  {journal} {Physical Review}\ }\textbf {\bibinfo {volume} {188}},\
  \bibinfo {pages} {1969} (\bibinfo {year} {1969})}\BibitemShut {NoStop}%
\bibitem [{\citenamefont {Muller}\ \emph {et~al.}(2007)\citenamefont {Muller},
  \citenamefont {Flagg}, \citenamefont {Bianucci}, \citenamefont {Wang},
  \citenamefont {Deppe}, \citenamefont {Ma}, \citenamefont {Zhang},
  \citenamefont {Salamo}, \citenamefont {Xiao},\ and\ \citenamefont
  {Shih}}]{muller_resonance_2007}%
  \BibitemOpen
  \bibfield  {author} {\bibinfo {author} {\bibfnamefont {A.}~\bibnamefont
  {Muller}}, \bibinfo {author} {\bibfnamefont {E.~B.}\ \bibnamefont {Flagg}},
  \bibinfo {author} {\bibfnamefont {P.}~\bibnamefont {Bianucci}}, \bibinfo
  {author} {\bibfnamefont {X.~Y.}\ \bibnamefont {Wang}}, \bibinfo {author}
  {\bibfnamefont {D.~G.}\ \bibnamefont {Deppe}}, \bibinfo {author}
  {\bibfnamefont {W.}~\bibnamefont {Ma}}, \bibinfo {author} {\bibfnamefont
  {J.}~\bibnamefont {Zhang}}, \bibinfo {author} {\bibfnamefont {G.~J.}\
  \bibnamefont {Salamo}}, \bibinfo {author} {\bibfnamefont {M.}~\bibnamefont
  {Xiao}}, \ and\ \bibinfo {author} {\bibfnamefont {C.~K.}\ \bibnamefont
  {Shih}},\ }\href {\doibase 10.1103/PhysRevLett.99.187402} {\bibfield
  {journal} {\bibinfo  {journal} {Physical Review Letters}\ }\textbf {\bibinfo
  {volume} {99}},\ \bibinfo {pages} {187402} (\bibinfo {year}
  {2007})}\BibitemShut {NoStop}%
\bibitem [{\citenamefont {Flagg}\ \emph {et~al.}(2009)\citenamefont {Flagg},
  \citenamefont {Muller}, \citenamefont {Robertson}, \citenamefont {Founta},
  \citenamefont {Deppe}, \citenamefont {Xiao}, \citenamefont {Ma},
  \citenamefont {Salamo},\ and\ \citenamefont {Shih}}]{flagg_resonantly_2009}%
  \BibitemOpen
  \bibfield  {author} {\bibinfo {author} {\bibfnamefont {E.~B.}\ \bibnamefont
  {Flagg}}, \bibinfo {author} {\bibfnamefont {A.}~\bibnamefont {Muller}},
  \bibinfo {author} {\bibfnamefont {J.~W.}\ \bibnamefont {Robertson}}, \bibinfo
  {author} {\bibfnamefont {S.}~\bibnamefont {Founta}}, \bibinfo {author}
  {\bibfnamefont {D.~G.}\ \bibnamefont {Deppe}}, \bibinfo {author}
  {\bibfnamefont {M.}~\bibnamefont {Xiao}}, \bibinfo {author} {\bibfnamefont
  {W.}~\bibnamefont {Ma}}, \bibinfo {author} {\bibfnamefont {G.~J.}\
  \bibnamefont {Salamo}}, \ and\ \bibinfo {author} {\bibfnamefont {C.~K.}\
  \bibnamefont {Shih}},\ }\href {\doibase 10.1038/nphys1184} {\bibfield
  {journal} {\bibinfo  {journal} {Nature Physics}\ }\textbf {\bibinfo {volume}
  {5}},\ \bibinfo {pages} {203} (\bibinfo {year} {2009})}\BibitemShut {NoStop}%
\bibitem [{\citenamefont {Ge}\ \emph {et~al.}(2013)\citenamefont {Ge},
  \citenamefont {Weiler}, \citenamefont {Ulhaq}, \citenamefont {Ulrich},
  \citenamefont {Jetter}, \citenamefont {Michler},\ and\ \citenamefont
  {Hughes}}]{ge_mollow_2013}%
  \BibitemOpen
  \bibfield  {author} {\bibinfo {author} {\bibfnamefont {R.-C.}\ \bibnamefont
  {Ge}}, \bibinfo {author} {\bibfnamefont {S.}~\bibnamefont {Weiler}}, \bibinfo
  {author} {\bibfnamefont {A.}~\bibnamefont {Ulhaq}}, \bibinfo {author}
  {\bibfnamefont {S.~M.}\ \bibnamefont {Ulrich}}, \bibinfo {author}
  {\bibfnamefont {M.}~\bibnamefont {Jetter}}, \bibinfo {author} {\bibfnamefont
  {P.}~\bibnamefont {Michler}}, \ and\ \bibinfo {author} {\bibfnamefont
  {S.}~\bibnamefont {Hughes}},\ }\href {\doibase 10.1364/OL.38.001691}
  {\bibfield  {journal} {\bibinfo  {journal} {Optics Letters}\ }\textbf
  {\bibinfo {volume} {38}},\ \bibinfo {pages} {1691} (\bibinfo {year}
  {2013})}\BibitemShut {NoStop}%
\bibitem [{\citenamefont {He}\ \emph {et~al.}(2013)\citenamefont {He},
  \citenamefont {He}, \citenamefont {Wei}, \citenamefont {Wu}, \citenamefont
  {Atatüre}, \citenamefont {Schneider}, \citenamefont {Höfling},
  \citenamefont {Kamp}, \citenamefont {Lu},\ and\ \citenamefont
  {Pan}}]{he_-demand_2013}%
  \BibitemOpen
  \bibfield  {author} {\bibinfo {author} {\bibfnamefont {Y.-M.}\ \bibnamefont
  {He}}, \bibinfo {author} {\bibfnamefont {Y.}~\bibnamefont {He}}, \bibinfo
  {author} {\bibfnamefont {Y.-J.}\ \bibnamefont {Wei}}, \bibinfo {author}
  {\bibfnamefont {D.}~\bibnamefont {Wu}}, \bibinfo {author} {\bibfnamefont
  {M.}~\bibnamefont {Atatüre}}, \bibinfo {author} {\bibfnamefont
  {C.}~\bibnamefont {Schneider}}, \bibinfo {author} {\bibfnamefont
  {S.}~\bibnamefont {Höfling}}, \bibinfo {author} {\bibfnamefont
  {M.}~\bibnamefont {Kamp}}, \bibinfo {author} {\bibfnamefont {C.-Y.}\
  \bibnamefont {Lu}}, \ and\ \bibinfo {author} {\bibfnamefont {J.-W.}\
  \bibnamefont {Pan}},\ }\href {\doibase 10.1038/nnano.2012.262} {\bibfield
  {journal} {\bibinfo  {journal} {Nature Nanotechnology}\ }\textbf {\bibinfo
  {volume} {8}},\ \bibinfo {pages} {213} (\bibinfo {year} {2013})}\BibitemShut
  {NoStop}%
\bibitem [{\citenamefont {Kuhlmann}\ \emph {et~al.}(2015)\citenamefont
  {Kuhlmann}, \citenamefont {Prechtel}, \citenamefont {Houel}, \citenamefont
  {Ludwig}, \citenamefont {Reuter}, \citenamefont {Wieck},\ and\ \citenamefont
  {Warburton}}]{kuhlmann_transform-limited_2015}%
  \BibitemOpen
  \bibfield  {author} {\bibinfo {author} {\bibfnamefont {A.~V.}\ \bibnamefont
  {Kuhlmann}}, \bibinfo {author} {\bibfnamefont {J.~H.}\ \bibnamefont
  {Prechtel}}, \bibinfo {author} {\bibfnamefont {J.}~\bibnamefont {Houel}},
  \bibinfo {author} {\bibfnamefont {A.}~\bibnamefont {Ludwig}}, \bibinfo
  {author} {\bibfnamefont {D.}~\bibnamefont {Reuter}}, \bibinfo {author}
  {\bibfnamefont {A.~D.}\ \bibnamefont {Wieck}}, \ and\ \bibinfo {author}
  {\bibfnamefont {R.~J.}\ \bibnamefont {Warburton}},\ }\href {\doibase
  10.1038/ncomms9204} {\bibfield  {journal} {\bibinfo  {journal} {Nature
  Communications}\ }\textbf {\bibinfo {volume} {6}},\ \bibinfo {pages} {8204}
  (\bibinfo {year} {2015})}\BibitemShut {NoStop}%
\bibitem [{\citenamefont {Kuhlmann}\ \emph
  {et~al.}(2013{\natexlab{a}})\citenamefont {Kuhlmann}, \citenamefont {Houel},
  \citenamefont {Ludwig}, \citenamefont {Greuter}, \citenamefont {Reuter},
  \citenamefont {Wieck}, \citenamefont {Poggio},\ and\ \citenamefont
  {Warburton}}]{kuhlmann_charge_2013}%
  \BibitemOpen
  \bibfield  {author} {\bibinfo {author} {\bibfnamefont {A.~V.}\ \bibnamefont
  {Kuhlmann}}, \bibinfo {author} {\bibfnamefont {J.}~\bibnamefont {Houel}},
  \bibinfo {author} {\bibfnamefont {A.}~\bibnamefont {Ludwig}}, \bibinfo
  {author} {\bibfnamefont {L.}~\bibnamefont {Greuter}}, \bibinfo {author}
  {\bibfnamefont {D.}~\bibnamefont {Reuter}}, \bibinfo {author} {\bibfnamefont
  {A.~D.}\ \bibnamefont {Wieck}}, \bibinfo {author} {\bibfnamefont
  {M.}~\bibnamefont {Poggio}}, \ and\ \bibinfo {author} {\bibfnamefont {R.~J.}\
  \bibnamefont {Warburton}},\ }\href {\doibase 10.1038/nphys2688} {\bibfield
  {journal} {\bibinfo  {journal} {Nature Physics}\ }\textbf {\bibinfo {volume}
  {9}},\ \bibinfo {pages} {570} (\bibinfo {year}
  {2013}{\natexlab{a}})}\BibitemShut {NoStop}%
\bibitem [{\citenamefont {Stanley}\ \emph {et~al.}(2014)\citenamefont
  {Stanley}, \citenamefont {Matthiesen}, \citenamefont {Hansom}, \citenamefont
  {Le~Gall}, \citenamefont {Schulte}, \citenamefont {Clarke},\ and\
  \citenamefont {Atatüre}}]{stanley_dynamics_2014}%
  \BibitemOpen
  \bibfield  {author} {\bibinfo {author} {\bibfnamefont {M.~J.}\ \bibnamefont
  {Stanley}}, \bibinfo {author} {\bibfnamefont {C.}~\bibnamefont {Matthiesen}},
  \bibinfo {author} {\bibfnamefont {J.}~\bibnamefont {Hansom}}, \bibinfo
  {author} {\bibfnamefont {C.}~\bibnamefont {Le~Gall}}, \bibinfo {author}
  {\bibfnamefont {C.~H.~H.}\ \bibnamefont {Schulte}}, \bibinfo {author}
  {\bibfnamefont {E.}~\bibnamefont {Clarke}}, \ and\ \bibinfo {author}
  {\bibfnamefont {M.}~\bibnamefont {Atatüre}},\ }\href {\doibase
  10.1103/PhysRevB.90.195305} {\bibfield  {journal} {\bibinfo  {journal}
  {Physical Review B}\ }\textbf {\bibinfo {volume} {90}},\ \bibinfo {pages}
  {195305} (\bibinfo {year} {2014})}\BibitemShut {NoStop}%
\bibitem [{\citenamefont {Chen}\ \emph {et~al.}(2016)\citenamefont {Chen},
  \citenamefont {Lander}, \citenamefont {Krowpman}, \citenamefont {Solomon},\
  and\ \citenamefont {Flagg}}]{chen_characterization_2016}%
  \BibitemOpen
  \bibfield  {author} {\bibinfo {author} {\bibfnamefont {D.}~\bibnamefont
  {Chen}}, \bibinfo {author} {\bibfnamefont {G.~R.}\ \bibnamefont {Lander}},
  \bibinfo {author} {\bibfnamefont {K.~S.}\ \bibnamefont {Krowpman}}, \bibinfo
  {author} {\bibfnamefont {G.~S.}\ \bibnamefont {Solomon}}, \ and\ \bibinfo
  {author} {\bibfnamefont {E.~B.}\ \bibnamefont {Flagg}},\ }\href {\doibase
  10.1103/PhysRevB.93.115307} {\bibfield  {journal} {\bibinfo  {journal}
  {Physical Review B}\ }\textbf {\bibinfo {volume} {93}},\ \bibinfo {pages}
  {115307} (\bibinfo {year} {2016})}\BibitemShut {NoStop}%
\bibitem [{\citenamefont {Vamivakas}\ \emph {et~al.}(2009)\citenamefont
  {Vamivakas}, \citenamefont {Zhao}, \citenamefont {Lu},\ and\ \citenamefont
  {Atatüre}}]{vamivakas_spin-resolved_2009}%
  \BibitemOpen
  \bibfield  {author} {\bibinfo {author} {\bibfnamefont {N.~A.}\ \bibnamefont
  {Vamivakas}}, \bibinfo {author} {\bibfnamefont {Y.}~\bibnamefont {Zhao}},
  \bibinfo {author} {\bibfnamefont {C.-Y.}\ \bibnamefont {Lu}}, \ and\ \bibinfo
  {author} {\bibfnamefont {M.}~\bibnamefont {Atatüre}},\ }\href {\doibase
  10.1038/nphys1182} {\bibfield  {journal} {\bibinfo  {journal} {Nature
  Physics}\ }\textbf {\bibinfo {volume} {5}},\ \bibinfo {pages} {198} (\bibinfo
  {year} {2009})}\BibitemShut {NoStop}%
\bibitem [{\citenamefont {Kuhlmann}\ \emph
  {et~al.}(2013{\natexlab{b}})\citenamefont {Kuhlmann}, \citenamefont {Houel},
  \citenamefont {Brunner}, \citenamefont {Ludwig}, \citenamefont {Reuter},
  \citenamefont {Wieck},\ and\ \citenamefont
  {Warburton}}]{kuhlmann_dark-field_2013}%
  \BibitemOpen
  \bibfield  {author} {\bibinfo {author} {\bibfnamefont {A.~V.}\ \bibnamefont
  {Kuhlmann}}, \bibinfo {author} {\bibfnamefont {J.}~\bibnamefont {Houel}},
  \bibinfo {author} {\bibfnamefont {D.}~\bibnamefont {Brunner}}, \bibinfo
  {author} {\bibfnamefont {A.}~\bibnamefont {Ludwig}}, \bibinfo {author}
  {\bibfnamefont {D.}~\bibnamefont {Reuter}}, \bibinfo {author} {\bibfnamefont
  {A.~D.}\ \bibnamefont {Wieck}}, \ and\ \bibinfo {author} {\bibfnamefont
  {R.~J.}\ \bibnamefont {Warburton}},\ }\href {\doibase 10.1063/1.4813879}
  {\bibfield  {journal} {\bibinfo  {journal} {Review of Scientific
  Instruments}\ }\textbf {\bibinfo {volume} {84}},\ \bibinfo {pages} {073905}
  (\bibinfo {year} {2013}{\natexlab{b}})}\BibitemShut {NoStop}%
\bibitem [{\citenamefont {He}\ \emph {et~al.}(2015)\citenamefont {He},
  \citenamefont {He}, \citenamefont {Liu}, \citenamefont {Wei}, \citenamefont
  {Ramírez}, \citenamefont {Atatüre}, \citenamefont {Schneider},
  \citenamefont {Kamp}, \citenamefont {Höfling}, \citenamefont {Lu},\ and\
  \citenamefont {Pan}}]{he_dynamically_2015}%
  \BibitemOpen
  \bibfield  {author} {\bibinfo {author} {\bibfnamefont {Y.}~\bibnamefont
  {He}}, \bibinfo {author} {\bibfnamefont {Y.-M.}\ \bibnamefont {He}}, \bibinfo
  {author} {\bibfnamefont {J.}~\bibnamefont {Liu}}, \bibinfo {author}
  {\bibfnamefont {Y.-J.}\ \bibnamefont {Wei}}, \bibinfo {author} {\bibfnamefont
  {H.}~\bibnamefont {Ramírez}}, \bibinfo {author} {\bibfnamefont
  {M.}~\bibnamefont {Atatüre}}, \bibinfo {author} {\bibfnamefont
  {C.}~\bibnamefont {Schneider}}, \bibinfo {author} {\bibfnamefont
  {M.}~\bibnamefont {Kamp}}, \bibinfo {author} {\bibfnamefont {S.}~\bibnamefont
  {Höfling}}, \bibinfo {author} {\bibfnamefont {C.-Y.}\ \bibnamefont {Lu}}, \
  and\ \bibinfo {author} {\bibfnamefont {J.-W.}\ \bibnamefont {Pan}},\ }\href
  {\doibase 10.1103/PhysRevLett.114.097402} {\bibfield  {journal} {\bibinfo
  {journal} {Physical Review Letters}\ }\textbf {\bibinfo {volume} {114}},\
  \bibinfo {pages} {097402} (\bibinfo {year} {2015})}\BibitemShut {NoStop}%
\bibitem [{\citenamefont {Gammon}\ \emph {et~al.}(1996)\citenamefont {Gammon},
  \citenamefont {Snow}, \citenamefont {Shanabrook}, \citenamefont {Katzer},\
  and\ \citenamefont {Park}}]{gammon_fine_1996}%
  \BibitemOpen
  \bibfield  {author} {\bibinfo {author} {\bibfnamefont {D.}~\bibnamefont
  {Gammon}}, \bibinfo {author} {\bibfnamefont {E.~S.}\ \bibnamefont {Snow}},
  \bibinfo {author} {\bibfnamefont {B.~V.}\ \bibnamefont {Shanabrook}},
  \bibinfo {author} {\bibfnamefont {D.~S.}\ \bibnamefont {Katzer}}, \ and\
  \bibinfo {author} {\bibfnamefont {D.}~\bibnamefont {Park}},\ }\href {\doibase
  10.1103/PhysRevLett.76.3005} {\bibfield  {journal} {\bibinfo  {journal}
  {Physical Review Letters}\ }\textbf {\bibinfo {volume} {76}},\ \bibinfo
  {pages} {3005} (\bibinfo {year} {1996})}\BibitemShut {NoStop}%
\bibitem [{\citenamefont {Ivchenko}(1997)}]{ivchenko_fine_1997}%
  \BibitemOpen
  \bibfield  {author} {\bibinfo {author} {\bibfnamefont {E.~L.}\ \bibnamefont
  {Ivchenko}},\ }\href {\doibase
  10.1002/1521-396X(199711)164:1<487::AID-PSSA487>3.0.CO;2-1} {\bibfield
  {journal} {\bibinfo  {journal} {physica status solidi (a)}\ }\textbf
  {\bibinfo {volume} {164}},\ \bibinfo {pages} {487} (\bibinfo {year}
  {1997})}\BibitemShut {NoStop}%
\bibitem [{\citenamefont {Bayer}\ \emph {et~al.}(2002)\citenamefont {Bayer},
  \citenamefont {Ortner}, \citenamefont {Stern}, \citenamefont {Kuther},
  \citenamefont {Gorbunov}, \citenamefont {Forchel}, \citenamefont {Hawrylak},
  \citenamefont {Fafard}, \citenamefont {Hinzer}, \citenamefont {Reinecke},
  \citenamefont {Walck}, \citenamefont {Reithmaier}, \citenamefont {Klopf},\
  and\ \citenamefont {Schäfer}}]{bayer_fine_2002}%
  \BibitemOpen
  \bibfield  {author} {\bibinfo {author} {\bibfnamefont {M.}~\bibnamefont
  {Bayer}}, \bibinfo {author} {\bibfnamefont {G.}~\bibnamefont {Ortner}},
  \bibinfo {author} {\bibfnamefont {O.}~\bibnamefont {Stern}}, \bibinfo
  {author} {\bibfnamefont {A.}~\bibnamefont {Kuther}}, \bibinfo {author}
  {\bibfnamefont {A.~A.}\ \bibnamefont {Gorbunov}}, \bibinfo {author}
  {\bibfnamefont {A.}~\bibnamefont {Forchel}}, \bibinfo {author} {\bibfnamefont
  {P.}~\bibnamefont {Hawrylak}}, \bibinfo {author} {\bibfnamefont
  {S.}~\bibnamefont {Fafard}}, \bibinfo {author} {\bibfnamefont
  {K.}~\bibnamefont {Hinzer}}, \bibinfo {author} {\bibfnamefont {T.~L.}\
  \bibnamefont {Reinecke}}, \bibinfo {author} {\bibfnamefont {S.~N.}\
  \bibnamefont {Walck}}, \bibinfo {author} {\bibfnamefont {J.~P.}\ \bibnamefont
  {Reithmaier}}, \bibinfo {author} {\bibfnamefont {F.}~\bibnamefont {Klopf}}, \
  and\ \bibinfo {author} {\bibfnamefont {F.}~\bibnamefont {Schäfer}},\ }\href
  {\doibase 10.1103/PhysRevB.65.195315} {\bibfield  {journal} {\bibinfo
  {journal} {Physical Review B}\ }\textbf {\bibinfo {volume} {65}},\ \bibinfo
  {pages} {195315} (\bibinfo {year} {2002})}\BibitemShut {NoStop}%
\bibitem [{\citenamefont {Tong}\ and\ \citenamefont
  {Wu}(2011)}]{tong_theory_2011}%
  \BibitemOpen
  \bibfield  {author} {\bibinfo {author} {\bibfnamefont {H.}~\bibnamefont
  {Tong}}\ and\ \bibinfo {author} {\bibfnamefont {M.~W.}\ \bibnamefont {Wu}},\
  }\href {\doibase 10.1103/PhysRevB.83.235323} {\bibfield  {journal} {\bibinfo
  {journal} {Physical Review B}\ }\textbf {\bibinfo {volume} {83}},\ \bibinfo
  {pages} {235323} (\bibinfo {year} {2011})}\BibitemShut {NoStop}%
\bibitem [{\citenamefont {Nguyen}\ \emph {et~al.}(2013)\citenamefont {Nguyen},
  \citenamefont {Sallen}, \citenamefont {Abbarchi}, \citenamefont {Ferreira},
  \citenamefont {Voisin}, \citenamefont {Roussignol}, \citenamefont
  {Cassabois},\ and\ \citenamefont
  {Diederichs}}]{nguyen_photoneutralization_2013}%
  \BibitemOpen
  \bibfield  {author} {\bibinfo {author} {\bibfnamefont {H.~S.}\ \bibnamefont
  {Nguyen}}, \bibinfo {author} {\bibfnamefont {G.}~\bibnamefont {Sallen}},
  \bibinfo {author} {\bibfnamefont {M.}~\bibnamefont {Abbarchi}}, \bibinfo
  {author} {\bibfnamefont {R.}~\bibnamefont {Ferreira}}, \bibinfo {author}
  {\bibfnamefont {C.}~\bibnamefont {Voisin}}, \bibinfo {author} {\bibfnamefont
  {P.}~\bibnamefont {Roussignol}}, \bibinfo {author} {\bibfnamefont
  {G.}~\bibnamefont {Cassabois}}, \ and\ \bibinfo {author} {\bibfnamefont
  {C.}~\bibnamefont {Diederichs}},\ }\href {\doibase
  10.1103/PhysRevB.87.115305} {\bibfield  {journal} {\bibinfo  {journal}
  {Physical Review B}\ }\textbf {\bibinfo {volume} {87}},\ \bibinfo {pages}
  {115305} (\bibinfo {year} {2013})}\BibitemShut {NoStop}%
\bibitem [{\citenamefont {Lu}\ and\ \citenamefont
  {Chipman}(1996)}]{lu_interpretation_1996}%
  \BibitemOpen
  \bibfield  {author} {\bibinfo {author} {\bibfnamefont {S.-Y.}\ \bibnamefont
  {Lu}}\ and\ \bibinfo {author} {\bibfnamefont {R.~A.}\ \bibnamefont
  {Chipman}},\ }\href {\doibase 10.1364/JOSAA.13.001106} {\bibfield  {journal}
  {\bibinfo  {journal} {Journal of the Optical Society of America A}\ }\textbf
  {\bibinfo {volume} {13}},\ \bibinfo {pages} {1106} (\bibinfo {year}
  {1996})}\BibitemShut {NoStop}%
\bibitem [{\citenamefont {Allen}\ and\ \citenamefont
  {Eberly}(1975)}]{allen_optical_1975}%
  \BibitemOpen
  \bibfield  {author} {\bibinfo {author} {\bibfnamefont {L.}~\bibnamefont
  {Allen}}\ and\ \bibinfo {author} {\bibfnamefont {J.~H.}\ \bibnamefont
  {Eberly}},\ }\href@noop {} {\emph {\bibinfo {title} {Optical {Resonance} and
  {Two} {Level} {Atoms}}}}\ (\bibinfo  {publisher} {Wiley, New York},\ \bibinfo
  {year} {1975})\BibitemShut {NoStop}%
\bibitem [{\citenamefont {Bonadeo}\ \emph {et~al.}(1998)\citenamefont
  {Bonadeo}, \citenamefont {Erland}, \citenamefont {Gammon}, \citenamefont
  {Park}, \citenamefont {Katzer},\ and\ \citenamefont
  {Steel}}]{bonadeo_coherent_1998}%
  \BibitemOpen
  \bibfield  {author} {\bibinfo {author} {\bibfnamefont {N.~H.}\ \bibnamefont
  {Bonadeo}}, \bibinfo {author} {\bibfnamefont {J.}~\bibnamefont {Erland}},
  \bibinfo {author} {\bibfnamefont {D.}~\bibnamefont {Gammon}}, \bibinfo
  {author} {\bibfnamefont {D.}~\bibnamefont {Park}}, \bibinfo {author}
  {\bibfnamefont {D.~S.}\ \bibnamefont {Katzer}}, \ and\ \bibinfo {author}
  {\bibfnamefont {D.~G.}\ \bibnamefont {Steel}},\ }\href {\doibase
  10.1126/science.282.5393.1473} {\bibfield  {journal} {\bibinfo  {journal}
  {Science}\ }\textbf {\bibinfo {volume} {282}},\ \bibinfo {pages} {1473}
  (\bibinfo {year} {1998})}\BibitemShut {NoStop}%
\bibitem [{\citenamefont {Unold}\ \emph {et~al.}(2004)\citenamefont {Unold},
  \citenamefont {Mueller}, \citenamefont {Lienau}, \citenamefont {Elsaesser},\
  and\ \citenamefont {Wieck}}]{unold_optical_2004}%
  \BibitemOpen
  \bibfield  {author} {\bibinfo {author} {\bibfnamefont {T.}~\bibnamefont
  {Unold}}, \bibinfo {author} {\bibfnamefont {K.}~\bibnamefont {Mueller}},
  \bibinfo {author} {\bibfnamefont {C.}~\bibnamefont {Lienau}}, \bibinfo
  {author} {\bibfnamefont {T.}~\bibnamefont {Elsaesser}}, \ and\ \bibinfo
  {author} {\bibfnamefont {A.~D.}\ \bibnamefont {Wieck}},\ }\href {\doibase
  10.1103/PhysRevLett.92.157401} {\bibfield  {journal} {\bibinfo  {journal}
  {Physical Review Letters}\ }\textbf {\bibinfo {volume} {92}},\ \bibinfo
  {pages} {157401} (\bibinfo {year} {2004})}\BibitemShut {NoStop}%
\bibitem [{\citenamefont {Muller}\ \emph {et~al.}(2008)\citenamefont {Muller},
  \citenamefont {Fang}, \citenamefont {Lawall},\ and\ \citenamefont
  {Solomon}}]{muller_emission_2008}%
  \BibitemOpen
  \bibfield  {author} {\bibinfo {author} {\bibfnamefont {A.}~\bibnamefont
  {Muller}}, \bibinfo {author} {\bibfnamefont {W.}~\bibnamefont {Fang}},
  \bibinfo {author} {\bibfnamefont {J.}~\bibnamefont {Lawall}}, \ and\ \bibinfo
  {author} {\bibfnamefont {G.~S.}\ \bibnamefont {Solomon}},\ }\href {\doibase
  10.1103/PhysRevLett.101.027401} {\bibfield  {journal} {\bibinfo  {journal}
  {Physical Review Letters}\ }\textbf {\bibinfo {volume} {101}},\ \bibinfo
  {pages} {027401} (\bibinfo {year} {2008})}\BibitemShut {NoStop}%
\end{thebibliography}%
\end{document}